\documentclass[conference]{IEEEtran}
\IEEEoverridecommandlockouts
\usepackage{cite}
\usepackage[utf8]{inputenc} 
\usepackage[T1]{fontenc}
\usepackage{amssymb,amsfonts}
\usepackage[cmex10]{amsmath}
\usepackage{algorithm}
\usepackage{algpseudocode}
\let\oldReturn\Return
\renewcommand{\Return}{\State\oldReturn}
\allowdisplaybreaks
\usepackage{graphicx}
\usepackage{textcomp}
\usepackage{xcolor}
\usepackage{pgfplots}
\usepackage{subcaption}
\usepackage[keeplastbox]{flushend}
\usepackage{tikz}

\usetikzlibrary{matrix, positioning, arrows}

\def\BibTeX{{\rm B\kern-.05em{\sc i\kern-.025em b}\kern-.08em
    T\kern-.1667em\lower.7ex\hbox{E}\kern-.125emX}}

\begin{document}

\title{Permutation Decoding of Polar Codes}

\author{
\IEEEauthorblockN{Mikhail Kamenev, Yulia Kameneva, Oleg Kurmaev, Alexey Maevskiy}
\IEEEauthorblockA{Moscow Research Center, Huawei Technologies Co., Ltd. \\
Moscow, Russia \\
Email: \{kamenev.mikhail1, kameneva.iuliia, oleg.kurmaev, maevskiy.alexey\}@huawei.com}
}

\maketitle

\begin{abstract}
A new permutation decoding approach for polar codes is presented. The complexity of the algorithm is similar to that of a successive cancellation list (SCL) decoder, while it can be implemented with the latency of a successive cancellation decoder. As opposed to the SCL algorithm, the sorting operation is not used in the proposed method.  It is shown that the error correction performance of the algorithm is similar to that of the SCL decoder for polar codes. Moreover, a new construction aiming to improve the error correction performance of polar codes under the proposed algorithm is presented.
\end{abstract}

\section{Introduction}
Polar codes have been shown to achieve the symmetric capacity of any binary-input discrete memoryless channel under a low-complex successive cancellation (SC) decoder \cite{Arikan}. However, the performance of the finite length polar codes under the SC decoder is quite poor. To solve the issue, a successive cancellation list (SCL) decoder has been proposed \cite{SCL}. It allows getting performance very close to that of maximum-likelihood decoding. It has been shown that if some additional information is used to select the correct codeword from the list, e.g. cyclic redundancy check (CRC), then it is possible to achieve the performance of the state-of-the-art low-density parity check codes.

The bottleneck of the SCL decoding algorithm with a large list size is sorting operation increasing the decoding latency \cite{MinSum}. One of the approaches to get rid of sorting, while keeping the error correction performance close to that of the SCL decoder, is an SC flip decoder \cite{SCFlip}. If the codeword returned by the algorithm does not pass the CRC, then the algorithm has several attempts to identify the first error bit and to flip it. Some approaches have been proposed recently aiming to enhance the identification of the first error bit index \cite{SCFlip2}. The procedures allow improving the performance of the SC algorithm while requiring the same amount of memory \cite{SCFlip}. 


In the paper, a new permutation decoding method for polar codes is proposed. Its error correction performance is similar to that of the SCL decoding method, while it can be implemented with the latency of the SC decoder and the hardware complexity $\mathcal{O}(Ln)$, where $n$ is the code length and $L$ is the list size. It is shown that the method can be used for decoding of polar codes optimized for the SCL algorithm. Also, an algorithm for the optimization of the frozen bits set for considered permutation decoding is presented.

The rest of the paper is organized as follows. Section II provides a general description of polar codes. In section III a new permutation decoding method for polar codes is presented. In section IV we propose a method for finding good permutations and an approach for the optimization of the frozen bits set for the proposed permutation decoding algorithm. Numerical results are presented in section V. We conclude the paper in section VI.

\section{Polar codes}
$(n, k)$ polar code \cite{Arikan} is a linear block code of length $n=2^m$, where $m$ is some positive integer, and dimension $k$ generated by $k$ rows $j_i \in \left\lbrace0, 1, \dots , n - 1\right\rbrace \setminus \mathcal{F}$, $0 \leq i < k$ of the matrix
\begin{equation}
\mathbf{A_m} = \begin{bmatrix}
    1 & 0 \\
    1 & 1 \\
\end{bmatrix}^{\otimes m},
\end{equation}
where $\mathbf{X}^{\otimes m}$ denotes $m$-times Kronecker product of the matrix $\mathbf{X}$ with itself. The set of frozen bits $\mathcal{F}$ is constructed as a set of indices $i$ maximizing  error correction performance of the code. For instance, Gaussian approximation (GA) for density evolution \cite{GA} can be used to generate polar codes having optimal error correction performance under the SC decoding algorithm in the binary-input additive white Gaussian noise (BI-AWGN) channel. 

Polar codes log likelihood ratio (LLR) based SCL decoding \cite{MinSum} can be efficiently implemented using the factor graph representation \cite{Arikan}. LLRs are calculated in a recursive manner using the following operations
\begin{subequations}
\begin{align}
f_{-}(x,y) &\triangleq \ln_{}\left(\frac{e^{x + y} + 1}{e^{x} + e^{y}}\right), \label{softXor} \\ 
f_{+}(x,y,u) &\triangleq \left(1 - 2u\right)x + y,  \label{softSum}
\end{align}
\end{subequations}
where $x$ and $y$ are real, while $u$ is binary. Instead of using (\ref{softXor}), we will follow the approach proposed in \cite{MinSum} and use the hardware-friendly approximation, namely
\begin{equation}
f_{-}\left(x,y\right) \approx \tilde{f_{-}}\left(x,y\right)\triangleq \textrm{sign}\left(x\right)\textrm{sign}\left(y\right)\textrm{min}\left\lbrace\left|x\right|, \left|y\right|\right\rbrace.
\end{equation}

 \begin{figure}[tbp]

 \begin{algorithmic}
 \renewcommand{\algorithmicrequire}{\textbf{Input:}}
 \renewcommand{\algorithmicensure}{\textbf{Output:}}
 \Require  A vector of LLRs $\mathbf{\hat{y}^l}$, a vector of bits $\mathbf{\hat{u}^0}$, a set of the frozen bits $\mathcal{F}$, an  index of outer code $g$, a layer index $l$, a permutation  $\pi$.
 \Ensure A decoded codeword metric $M$, a vector of bits $\mathbf{\hat{u}^{l}}$.
 \Function{SC}{$\mathbf{\hat{y}^l}, \mathbf{\hat{u}^0}, \mathcal{F},  g, l, \pi$}
\State Set $\mathbf{\hat{u}^l}$ to be all zeros vector of size $2^l$
 \If {$l = 0$}
 \If {$\pi\left(g\right) \in \mathcal{F}$}
 \State $\mathbf{\hat{u}^0}[g] \gets 0$, $\mathbf{\hat{u}^l}[0]\gets 0$
 \State $M \gets \textrm{min}\left\lbrace0, \mathbf{\hat{y}^l}[0]\right\rbrace$
 \Else
 \If {$\mathbf{\hat{y}^l}[0] \leq 0$}
 \State $\mathbf{\hat{u}^0}[g]\gets 1$, $\mathbf{\hat{u}^l}[0]\gets 1$
 \Else
 \State $\mathbf{\hat{u}^0}[g] \gets 0$, $\mathbf{\hat{u}^l}[0]\gets 0$
 \EndIf
 \State  $M \gets 0$
 \EndIf
 \Return $M$, $\mathbf{\hat{u}^{l}}$
 \EndIf
 \State Set $\mathbf{\hat{y}^{l-1}}$ to be all zeros vector of size $2^{l-1}$\
 \For {$i = 0$ to $2^{l-1}-1$}
 \State $\mathbf{\hat{y}^{l-1}}[i] \gets \tilde{f_{-}}\left(\mathbf{\hat{y}^{l}}[i],\mathbf{\hat{y}^{l}}[i+2^{l-1}]\right)$
 \EndFor
  \State $M, \mathbf{\hat{u}^{l-1}}  \gets \texttt{SC}\left(\mathbf{\hat{y}^{l-1}},\mathbf{\hat{u}^0},\mathcal{F},  2g, l - 1, \pi \right)$
 \For {$i = 0$ to $2^{l-1}-1$}
 \State $\mathbf{\hat{u}^{l}}[i] \gets \mathbf{\hat{u}^{l-1}}[i]$
 \State $\mathbf{\hat{y}^{l-1}}[i] \gets f_{+}\left(\mathbf{\hat{y}^{l}}[i],\mathbf{\hat{y}^{l}}[ i + 2^{l-1}], \mathbf{\hat{u}^{l-1}}[i] \right)$
 \EndFor
\State $M^\prime,\mathbf{\hat{u}^{l-1}} \gets \texttt{SC}\left(\mathbf{\hat{y}^{l-1}},\mathbf{\hat{u}^0},\mathcal{F},  2g + 1, l - 1, \pi\right)$
 \State $M \gets M + M^\prime$
 \For {$i = 0$ to $2^{l-1}-1$}
 \State $\mathbf{\hat{u}^{l}}[i] \gets \mathbf{\hat{u}^{l}}[i] \oplus \mathbf{\hat{u}^{l-1}}[i]$
 \State $\mathbf{\hat{u}^{l}}[i+ 2^{l-1}] \gets \mathbf{\hat{u}^{l-1}}[i]$
 \EndFor
 \Return $M$, $\mathbf{\hat{u}^{l}}$ 
 \EndFunction 
 \end{algorithmic} 
  \caption{Recursive calculations used in the SC algorithm.}
 \label{RecursivelyCalcSC}
 \end{figure}

\section{Permutation decoding algorithm}
Recall that a permutation group of the code contains permutations of the code positions that does not change the set of codewords, i.e. transform any codeword of the code to another or the same codeword. For instance, Reed-Muller codes have the permutation group, which is isomorphic to the whole affine group $GA(m)$ \cite[Sec.~13.9]{Sloan}. Unfortunately, the permutation group of an arbitrary polar code have no explicit construction \cite{PermGroupPolar}.  Here, for simplicity, we will consider only $m!$ factor graph layers permutations $\pi^l:\left(0,1, \dots, m-1 \right) \rightarrow \left(\pi^l\left(0\right), \pi^l\left(1\right), \dots, \pi^l\left(m-1\right) \right)$ \cite{Perm}. Let $\pi$ be the corresponding bit indices permutation. 

The set of frozen bits may be changed after applying a factor graph layers permutation. In such a case the permutation does not belong to the permutation group, but the code remains polar and can be decoded using the SC algorithm \cite{PermSC}. Moreover, such kind of permutation can lead to a significant error correction performance degradation. Thus, the joint optimization of the frozen bits set and the permutations set is required. Also, it is possible to optimize the permutations set for the fixed frozen bits set.

Here the following permutation decoding approach is proposed. Firstly, a permutations set containing $L$ factor graph layers permutations is generated.  Then, the SC algorithm process $L$ permuted versions of the received channel LLRs, and return $L$ decoded codewords with the corresponding metrics. Finally, the codeword with the best metric is returned as the output of the algorithm. The formal description of the proposed permutation decoding method with the LLR based metric is presented in Figs. \ref{RecursivelyCalcSC} -- \ref{PermutationDecodingAlgorithm}. An example of the factor graph layers permutation and its effect on the SC decoding algorithm is depicted in Fig. \ref{FactorGraphPermutationExample}.

The latency and the hardware complexity of the proposed permutation algorithm can be estimated using that of the SC decoder. The latency of the SC decoder equals $\mathcal{O}(n\log{}n)$, while the hardware complexity equals $\mathcal{O}(n)$ \cite{SCHardware}, where $n$ is the code length. Observe that all instances of the SC decoder are independent and can be run in parallel. Thus, the latency of a parallel implementation equals that of the SC decoder, while the hardware complexity equals that of $L$ independent SC decoders, namely $\mathcal{O}(Ln)$. To minimize the hardware complexity of the proposed decoder, it is possible to run instances of the SC decoders one after another. In such a case, the hardware complexity of the considered permutation decoding method equals that of the SC decoder, namely $\mathcal{O}(n)$, while the latency equals $\mathcal{O}(Ln\log{}n)$. Also, the considered permutation decoding algorithm benefits from different hardware improvements of the SC decoder.

The main benefit of the proposed method is that it does not use a sorting operation. Thus it is more feasible for a hardware implementation than the SCL algorithm. However, the presented permutation decoder can return less than $L$ unique codewords, limiting the error correction performance in the case when the CRC is used to select the correct codeword from the list. Nevertheless, the fact can be used to significantly decrease the number of calculations, with the error correction performance degradation being negligible \cite{RMPerm}.

\section{Permutations and frozen bits sets optimization}

\subsection{Permutations set optimization for the fixed frozen bits set}
It has been observed that an arbitrary factor graph layers permutation can lead to a significant error correction performance degradation due to the frozen bits set change \cite{PermSC}. Thus, the permutations set should be optimized to maximize the error correction performance of the decoder. Since for the code of length $n = 2^m$ there are $m!$ possible permutations, it is impossible to use simulations to find the best permutations set for a large $n$. So, we propose a sub-optimal approach based on optimization of a lower bound for the error correction performance of the considered permutation decoder.

Let $X_{\pi_i}$ be the event that the SC decoder returns an incorrect codeword in the case the permutation $\pi_i$ is applied. Then the block error probability $P$ of the proposed permutation decoding algorithm with list size $L$ can be estimated as
\begin{equation}
\begin{aligned}
P &= P\left(\bigcap\limits_{i = 0}^{L-1}X_{\pi_i} \right) \\
&= P\left(X_{\pi_0}\right)P\left(X_{\pi_1}|X_{\pi_0}\right)\dots P\left(X_{\pi_{L-1}}|X_{\pi_0} \dots X_{\pi_{L-2}}\right).
\end{aligned}
\end{equation}

\begin{figure}[tbp]
 \begin{algorithmic}
 \renewcommand{\algorithmicrequire}{\textbf{Input:}}
 \renewcommand{\algorithmicensure}{\textbf{Output:}}
 \Require A code length $n$, a set of the frozen bits $\mathcal{F}$, a vector of received channel LLRs $\mathbf{\hat{y}}$, a set of permutations  $\mathcal{P}$ of size $L$.
 \Ensure  A vector of decoded bits $\mathbf{\hat{u}}$, a decoded codeword metric $M$.
 \Function{PermDecoding}{$n,\mathcal{F}, \mathbf{\hat{y}}, \mathcal{P}$}
 \State $M \gets -\infty$
 \State $m = \log_{2}{n}$
 \State Set $\mathbf{\hat{u}}$ to be all zeros vector of size $n$
 \ForAll {$\pi \in \mathcal{P}$}
 \State Set $\mathbf{\hat{u}^0}$ to be all zeros vector of size $n$
 \State $M^\prime, \mathbf{\hat{u}^m} \gets \texttt{SC}\left(\pi\left(\mathbf{\hat{y}}\right), \mathbf{\hat{u}^0}, \mathcal{F},  0, m, \pi\right)$
 \If {$M^\prime > M$}
 \State $M \gets M^\prime$
 \State $\mathbf{\hat{u}} \gets \pi^{-1}\left(\mathbf{\hat{u}^0}\right)$
 \EndIf
 \EndFor
 \Return $\mathbf{\hat{u}}, M$ 
 \EndFunction
 \end{algorithmic} 
 \caption{Permutation decoding algorithm.}
 \label{PermutationDecodingAlgorithm}
 \end{figure}

Since the conditional probability $P\left(X_{\pi_{j}}|X_{\pi_0} \dots X_{\pi_{j-1}}\right)$ is hard to evaluate, we will consider a lower bound for $P$, namely
\begin{equation}
P \geq \prod_{i = 0}^{L-1}\left(P\left(X_{\pi_i}\right)\right).
\label{pscl_lower_bound}
\end{equation}

$P\left(X_{\pi_i}\right)$ can be efficiently calculated using GA for density evolution. The algorithm can be used to estimate the error probability of the synthetic bits subchannels $\hat{P_i}$\cite{GA}, where $i$ is the channel index. Knowing all $\hat{P_i}$, the block error probability in the case the permutation $\pi$ is applied can be approximated as 
\begin{equation}
P\left(X_\pi\right) \approx 1 - \prod_{i \in \mathcal{I}}\left(1-\hat{P}_{\pi\left(i\right)}\right),
\label{bler_approximation}
\end{equation}
where $\mathcal{I} = \left\lbrace 0, 1, \dots, n - 1\right\rbrace \setminus \mathcal{F}$ \cite{GA}.
So, (\ref{pscl_lower_bound}) can be evaluated using the output of a single launch of GA for density evolution, and $L$ block error probability approximations (\ref{bler_approximation}).

Based on (\ref{pscl_lower_bound}), a permutations set $\mathcal{P}$ for the fixed frozen bits set can be optimized as follows. Firstly, the bit error probability is calculated for each synthetic subchannel. Then, the block error probability is calculated for each layers permutation. Finally, layers permutations are sorted by the corresponding block error probability in ascending order, and $L$ permutations with the lowest block error probability are selected.

Although (\ref{pscl_lower_bound}) allows getting a rough estimate of the block error probability of the considered permutation decoder, it does not take into account that two permutations having low block error probability can correct the same error patterns. This fact significantly affects the error correction performance of the permutation decoder optimized by the proposed algorithm. It has been observed that the layers permutations with the lowest block error probability are similar to each other in terms of Hamming distance. So, it is assumed that the layers permutations with the small Hamming distance correct a large number of identical error patterns. In order to solve the issue, it is proposed to select layers permutations having large enough Hamming distance to each other. As will be shown later, this limitation allows significantly improve the error correction performance of the proposed decoder.

\subsection{Frozen bits set optimization for permutation decoding}
Recall that Reed-Muller codes have the permutation group, which is isomorphic to the whole affine group $GA(m)$ \cite[Sec.~13.9]{Sloan}. Thus, any permutations set is optimal from (\ref{pscl_lower_bound}) point of view. Moreover, it has been observed that the error correction performance of the considered algorithm is similar to that of the SCL in the case of Reed-Muller codes decoding \cite{RMPerm}. Based on the observation, construction of polar codes having known automorphism group is proposed. The main idea of the approach is to select a set of layers that can be permuted randomly without the error correction performance degradation under SC decoding. Thus, the permutations set is constructed as a subset of such permutations. Nevertheless, one needs to construct a frozen bits set having both good error correction performance under SC decoding and a large enough automorphism group.

While optimizing $\left(n,k\right)$ polar code using GA for density evolution, one needs to choose $k$ information bits corresponding to synthetic subchannels having the largest capacity. In order to preserve the automorphism group, it is necessary to split bits indices into disjoint sets $G_0, \dots G_{t-1}$ in such a way, that 
\begin{equation}
\forall \pi \in \hat{\mathcal{P}}\forall i \in \left[0, t-1\right] \forall j \in G_i, \pi(j) \in G_i,
\end{equation}
where $\hat{\mathcal{P}}$ is a subset of permutations, which are used for decoding. We refer $\hat{\mathcal{P}}$ as a set of all available permutations. So, the set of frozen bits is constructed as a set of sets $G_0, \dots G_{t-1}$ having the lowest synthetic subchannels capacity. This construction guarantees that any permutation $\pi \in \hat{\mathcal{P}}$ can be efficiently used for the considered permutation decoding method without additional optimization.

\textit{Example:} Consider a polar code of length 32. Let a set of all available permutations be constructed as the subset of factor graph layers permutations $\left\lbrace\pi^l : \pi^l\left(0\right) = 0, \pi^l\left(1\right) = 1\right\rbrace$. In such a case, bits indices can be splitted as 
\begin{equation}
\begin{aligned}
\{
& \left\lbrace31\right\rbrace,
\left\lbrace15,23,27\right\rbrace,
\left\lbrace29\right\rbrace,
\left\lbrace30\right\rbrace,
\left\lbrace7,11,19\right\rbrace,
\left\lbrace28\right\rbrace, \\
& \left\lbrace13,21,25\right\rbrace,
\left\lbrace14,22,26\right\rbrace,
\left\lbrace3\right\rbrace,
\left\lbrace12,20,24\right\rbrace,
\left\lbrace5,9,17\right\rbrace, \\
& \left\lbrace6,10,18\right\rbrace,
\left\lbrace4,8,16\right\rbrace,
\left\lbrace1\right\rbrace,
\left\lbrace2\right\rbrace,
\left\lbrace0\right\rbrace
\}.
\end{aligned}
\label{splitting_example}
\end{equation}
It can be easily seen that a code of any dimension can be constructed using splitting (\ref{splitting_example}).

Based on a splitting, the frozen bits set construction task can be solved as a knapsack problem, with weights being equal to the cardinality of the set. Value of each set can be calculated as the maximum bit error probability of synthetic bit subchannel corresponding to an index in the set. Although it is not guaranteed that it is possible to construct a polar code of a given dimension $k$ in such a manner, one can construct a polar code of a dimension $k^\prime > k$ and remove some bit indices from the worst value set. In such a case, permutations from the set $\hat{\mathcal{P}}$ can lead to the frozen bits set change, but the block error probability for each permutation is upper bounded by the block error probability of the code with dimension $k^\prime$.
\begin{figure}[t]
\centering
\footnotesize
\begin{subfigure}[]{0.48\textwidth}
	\centering
	\begin{tikzpicture}[inner sep=0.5, >=stealth , scale=0.049, auto , bend angle = 25, ]
	\newcommand{\cc}{red!0}
	\newcommand{\cs}{$ + $}
	\newcommand{\vc}{black!255}
	\newcommand{\vs}{$ \phantom{\cdot} $}
	\newcommand{\dx}{20}
	\newcommand{\dxl}{10}
	\foreach \i in {0, 1,..., 7}
	\node at (-5, 70 - 10 * \i) [circle, draw, fill = \vc] (U\i) {};	
	\foreach \i in {0, 1,..., 3}
	\node at (1 * \dx, 70 - 20 * \i) [circle, draw, fill = \cc] (C0\i) {\cs};	
	\foreach \i in {0, 1,..., 3}
	\node at (1 * \dx, 60 - 20 * \i) [circle, draw, fill = \vc] (V0\i) {\vs};	
	\foreach \i in {0, 1,..., 3}
	\path [ - ,every node/.style = { font = \footnotesize } , outer sep = -3.0 ] (C0\i) edge node {} (V0\i);
	\foreach \i in {0, 2}	
	\node at (2 * \dx + \dxl, 70 - 20 * \i) [circle, draw, fill = \cc] (C1\i) {\cs};
	\foreach \i in {0, 2}	
	\node at (2 * \dx + \dxl, 50 - 20 * \i) [circle, draw, fill = \vc] (V1\i) {\vs};
	\foreach \i in {0, 2}		
	\path [ - ,every node/.style = { font = \footnotesize } , outer sep = -3.0 ] (C1\i) edge node {} (V1\i);
	\foreach \i in {1, 3}	
	\node at (2 * \dx, 80 - 20 * \i) [circle, draw, fill = \cc] (C1\i) {\cs};
	\foreach \i in {1, 3}	
	\node at (2 * \dx, 60 - 20 * \i) [circle, draw, fill = \vc] (V1\i) {\vs};
	\foreach \i in {1, 3}		
	\path [ - ,every node/.style = { font = \footnotesize } , outer sep = -3.0 ] (C1\i) edge node {} (V1\i);
	\node at (3 * \dx + \dxl * 4, 70) [circle, draw, fill = \cc] (C30) {\cs};
	\node at (3 * \dx + \dxl * 4, 30) [circle, draw, fill = \vc] (V30) {\vs};
	\path [ - ,every node/.style = { font = \footnotesize } , outer sep = -3.0 ] (C30) edge node {} (V30);
	\node at (3 * \dx + \dxl * 3, 60) [circle, draw, fill = \cc] (C31) {\cs};
	\node at (3 * \dx + \dxl * 3, 20) [circle, draw, fill = \vc] (V31) {\vs};
	\path [ - ,every node/.style = { font = \footnotesize } , outer sep = -3.0 ] (C31) edge node {} (V31);
	\node at (3 * \dx + \dxl * 2, 50) [circle, draw, fill = \cc] (C32) {\cs};
	\node at (3 * \dx + \dxl * 2, 10) [circle, draw, fill = \vc] (V32) {\vs};
	\path [ - ,every node/.style = { font = \footnotesize } , outer sep = -3.0 ] (C32) edge node {} (V32);
	\node at (3 * \dx + \dxl * 1, 40) [circle, draw, fill = \cc] (C33) {\cs};
	\node at (3 * \dx + \dxl * 1, 0) [circle, draw, fill = \vc] (V33) {\vs};
	\path [ - ,every node/.style = { font = \footnotesize } , outer sep = -3.0 ] (C33) edge node {} (V33);
	\foreach \i in {0, 1,..., 7}
	\node at (115, 70 - 10 * \i) [circle, draw, fill = \vc] (X\i) {};
	\path [ - ,every node/.style = { font = \footnotesize } , outer sep = -3.0 ]
	(U0) edge node {} (C00) (C00) edge node {} (C10) (C10) edge node {} (C30) (C30) edge node {} (X0)
	(U1) edge node {} (V00) (V00) edge node {} (C11) (C11) edge node {} (C31) (C31) edge node {} (X1)
	(U2) edge node {} (C01) (C01) edge node {} (V10) (V10) edge node {} (C32) (C32) edge node {} (X2)
	(U3) edge node {} (V01) (V01) edge node {} (V11) (V11) edge node {} (C33) (C33) edge node {} (X3)
	(U4) edge node {} (C02) (C02) edge node {} (C12) (C12) edge node {} (V30) (V30) edge node {} (X4)
	(U5) edge node {} (V02) (V02) edge node {} (C13) (C13) edge node {} (V31) (V31) edge node {} (X5)
	(U6) edge node {} (C03) (C03) edge node {} (V12) (V12) edge node {} (V32) (V32) edge node {} (X6)
	(U7) edge node {} (V03) (V03) edge node {} (V13) (V13) edge node {} (V33) (V33) edge node {} (X7);
	\node at (5, 75) {0.32};
	\node at (5, 65) {-1.33};
	\node at (5, 55) {-0.99};
	\node at (5, 45) {\textcolor[RGB]{139,0,0}{-0.14}};
	\node at (5, 35) {-1.16};
	\node at (5, 25) {1.49};
	\node at (5, 15) {-3.27};
	\node at (5, 5) {-13.49};
	
	\node at (110, 75) {-3.42};
	\node at (110, 65) {2.97};
	\node at (110, 55) {3.16};
	\node at (110, 45) {1.45};
	\node at (110, 35) {1.01};
	\node at (110, 25) {0.32};
	\node at (110, 15) {2.00};
	\node at (110, 5) {-6.12};

	\node at (20, -8) {0};
	\node at (47, -8) {1};
	\node at (87, -8) {2};
	
	\node at (122, 70) {0};
	\node at (122, 60) {1};
	\node at (122, 50) {2};
	\node at (122, 40) {3};
	\node at (122, 30) {4};
	\node at (122, 20) {5};
	\node at (122, 10) {6};
	\node at (122, 0) {7};
	
	\node at (-10, 70) {0};
	\node at (-10, 60) {1};
	\node at (-10, 50) {2};
	\node at (-10, 40) {3};
	\node at (-10, 30) {4};
	\node at (-10, 20) {5};
	\node at (-10, 10) {6};
	\node at (-10, 0) {7};

	\end{tikzpicture}
	\caption{}
	\label{PlainSC}
\end{subfigure}
\begin{subfigure}[]{0.48\textwidth}
	\centering
	\begin{tikzpicture}[inner sep=0.5, >=stealth , scale=0.049, auto , bend angle = 25, ]
	\newcommand{\cc}{red!0}
	\newcommand{\cs}{$ + $}
	\newcommand{\vc}{black!255}
	\newcommand{\vs}{$ \phantom{\cdot} $}
	\newcommand{\dx}{20}
	\newcommand{\dxl}{10}
	\foreach \i in {0, 1,..., 7}
	\node at (-5, 70 - 10 * \i) [circle, draw, fill = \vc] (U\i) {};	
	\foreach \i in {0, 1,..., 7}
	\coordinate  (UP\i) at (10, 70 - 10 * \i);	
	\foreach \i in {0, 1,..., 3}
	\node at (1 * \dx, 70 - 20 * \i) [circle, draw, fill = \cc] (C0\i) {\cs};	
	\foreach \i in {0, 1,..., 3}
	\node at (1 * \dx, 60 - 20 * \i) [circle, draw, fill = \vc] (V0\i) {\vs};	
	\foreach \i in {0, 1,..., 3}
	\path [ - ,every node/.style = { font = \footnotesize, outer sep = -3.0  }, outer sep = -3.0 ] (C0\i) edge node {} (V0\i);
	\foreach \i in {0, 2}	
	\node at (2 * \dx + \dxl, 70 - 20 * \i) [circle, draw, fill = \cc] (C1\i) {\cs};
	\foreach \i in {0, 2}	
	\node at (2 * \dx + \dxl, 50 - 20 * \i) [circle, draw, fill = \vc] (V1\i) {\vs};
	\foreach \i in {0, 2}		
	\path [ - ,every node/.style = { font = \footnotesize } , outer sep = -3.0 ] (C1\i) edge node {} (V1\i);
	\foreach \i in {1, 3}	
	\node at (2 * \dx, 80 - 20 * \i) [circle, draw, fill = \cc] (C1\i) {\cs};
	\foreach \i in {1, 3}	
	\node at (2 * \dx, 60 - 20 * \i) [circle, draw, fill = \vc] (V1\i) {\vs};
	\foreach \i in {1, 3}		
	\path [ - ,every node/.style = { font = \footnotesize } , outer sep = -3.0 ] (C1\i) edge node {} (V1\i);
	\node at (3 * \dx + \dxl * 4, 70) [circle, draw, fill = \cc] (C30) {\cs};
	\node at (3 * \dx + \dxl * 4, 30) [circle, draw, fill = \vc] (V30) {\vs};
	\path [ - ,every node/.style = { font = \footnotesize } , outer sep = -3.0 ] (C30) edge node {} (V30);
	\node at (3 * \dx + \dxl * 3, 60) [circle, draw, fill = \cc] (C31) {\cs};
	\node at (3 * \dx + \dxl * 3, 20) [circle, draw, fill = \vc] (V31) {\vs};
	\path [ - ,every node/.style = { font = \footnotesize } , outer sep = -3.0 ] (C31) edge node {} (V31);
	\node at (3 * \dx + \dxl * 2, 50) [circle, draw, fill = \cc] (C32) {\cs};
	\node at (3 * \dx + \dxl * 2, 10) [circle, draw, fill = \vc] (V32) {\vs};
	\path [ - ,every node/.style = { font = \footnotesize } , outer sep = -3.0 ] (C32) edge node {} (V32);
	\node at (3 * \dx + \dxl * 1, 40) [circle, draw, fill = \cc] (C33) {\cs};
	\node at (3 * \dx + \dxl * 1, 0) [circle, draw, fill = \vc] (V33) {\vs};
	\path [ - ,every node/.style = { font = \footnotesize } , outer sep = -3.0 ] (C33) edge node {} (V33);
	\foreach \i in {0, 1,..., 7}
	\node at (115, 70 - 10 * \i) [circle, draw, fill = \vc] (X\i) {};
	\path [ - ,every node/.style = { font = \footnotesize } , outer sep = -3.0 ]
	(U0) edge node {} (UP0) (UP0) edge node {} (C00) (C00) edge node {} (C10) (C10) edge node {} (C30) (C30) edge node {} (X0)
	(U1) edge node {} (UP1) (UP1) edge node {} (V00) (V00) edge node {} (C11) (C11) edge node {} (C31) (C31) edge node {} (X1)
	(U2) edge node {} (UP2) (UP4) edge node {} (C01) (C01) edge node {} (V10) (V10) edge node {} (C32) (C32) edge node {} (X2)
	(U3) edge node {} (UP3) (UP5) edge node {} (V01) (V01) edge node {} (V11) (V11) edge node {} (C33) (C33) edge node {} (X3)
	(U4) edge node {} (UP4) (UP2) edge node {} (C02) (C02) edge node {} (C12) (C12) edge node {} (V30) (V30) edge node {} (X4)
	(U5) edge node {} (UP5) (UP3) edge node {} (V02) (V02) edge node {} (C13) (C13) edge node {} (V31) (V31) edge node {} (X5)
	(U6) edge node {} (UP6) (UP6) edge node {} (C03) (C03) edge node {} (V12) (V12) edge node {} (V32) (V32) edge node {} (X6)
	(U7) edge node {} (UP7) (UP7) edge node {} (V03) (V03) edge node {} (V13) (V13) edge node {} (V33) (V33) edge node {} (X7);
	\node at (5, 75) {0.32};
	\node at (5, 65) {-1.33};
	\node at (5, 55) {0.99};
	\node at (5, 45) {2.51};
	\node at (5, 35) {-1.13};
	\node at (5, 25) {-1.02};
	\node at (5, 15) {-7.57};
	\node at (5, 5) {-15.53};
	
	\node at (110, 75) {-3.42};
	\node at (110, 65) {2.97};
	\node at (110, 55) {1.01};
	\node at (110, 45) {0.32};
	\node at (110, 35) {3.16};
	\node at (110, 25) {1.45};
	\node at (110, 15) {2.00};
	\node at (110, 5) {-6.12};

	\node at (20, -8) {0};
	\node at (47, -8) {1};
	\node at (87, -8) {2};
	
	\node at (122, 70) {0};
	\node at (122, 60) {1};
	\node at (122, 50) {4};
	\node at (122, 40) {5};
	\node at (122, 30) {2};
	\node at (122, 20) {3};
	\node at (122, 10) {6};
	\node at (122, 0) {7};
	
	\node at (-10, 70) {0};
	\node at (-10, 60) {1};
	\node at (-10, 50) {2};
	\node at (-10, 40) {3};
	\node at (-10, 30) {4};
	\node at (-10, 20) {5};
	\node at (-10, 10) {6};
	\node at (-10, 0) {7};
	\end{tikzpicture}
	\caption{}
	\label{LLRsPermutation}
\end{subfigure}
\begin{subfigure}[]{0.48\textwidth}
	\centering
	\begin{tikzpicture}[inner sep=0.5, >=stealth , scale=0.049, auto , bend angle = 25, ]
	\newcommand{\cc}{red!0}
	\newcommand{\cs}{$ + $}
	\newcommand{\vc}{black!255}
	\newcommand{\vs}{$ \phantom{\cdot} $}
	\newcommand{\dx}{20}
	\newcommand{\dxl}{10}
	\foreach \i in {0, 1,..., 7}
	\node at (-5, 70 - 10 * \i) [circle, draw, fill = \vc] (U\i) {};	
	
	\foreach \i in {0, 1,..., 3}
	\node at (1 * \dx, 70 - 20 * \i) [circle, draw, fill = \cc] (C0\i) {\cs};	
	\foreach \i in {0, 1,..., 3}
	\node at (1 * \dx, 60 - 20 * \i) [circle, draw, fill = \vc] (V0\i) {\vs};	
	\foreach \i in {0, 1,..., 3}
	\path [ - ,every node/.style = { font = \footnotesize } , outer sep = -3.0 ] (C0\i) edge node {} (V0\i);
	\foreach \i in {0, 2}	
	\node at (3 * \dx + \dxl * 4, 70 - 20 * \i) [circle, draw, fill = \cc] (C3\i) {\cs};
	\foreach \i in {0, 2}	
	\node at (3 * \dx + \dxl * 4, 50 - 20 * \i) [circle, draw, fill = \vc] (V3\i) {\vs};
	\foreach \i in {0, 2}		
	\path [ - ,every node/.style = { font = \footnotesize } , outer sep = -3.0 ] (C3\i) edge node {} (V3\i);
	\foreach \i in {1, 3}	
	\node at (3 * \dx + \dxl * 3, 80 - 20 * \i) [circle, draw, fill = \cc] (C3\i) {\cs};
	\foreach \i in {1, 3}	
	\node at (3 * \dx + \dxl * 3, 60 - 20 * \i) [circle, draw, fill = \vc] (V3\i) {\vs};
	\foreach \i in {1, 3}		
	\path [ - ,every node/.style = { font = \footnotesize } , outer sep = -3.0 ] (C3\i) edge node {} (V3\i);
	\node at (2 * \dx + \dxl * 3, 70) [circle, draw, fill = \cc] (C10) {\cs};
	\node at (2 * \dx + \dxl * 3, 30) [circle, draw, fill = \vc] (V10) {\vs};
	\path [ - ,every node/.style = { font = \footnotesize } , outer sep = -3.0 ] (C10) edge node {} (V10);
	\node at (2 * \dx + \dxl * 2, 60) [circle, draw, fill = \cc] (C11) {\cs};
	\node at (2 * \dx + \dxl * 2, 20) [circle, draw, fill = \vc] (V11) {\vs};
	\path [ - ,every node/.style = { font = \footnotesize } , outer sep = -3.0 ] (C11) edge node {} (V11);
	\node at (2 * \dx + \dxl * 1, 50) [circle, draw, fill = \cc] (C12) {\cs};
	\node at (2 * \dx + \dxl * 1, 10) [circle, draw, fill = \vc] (V12) {\vs};
	\path [ - ,every node/.style = { font = \footnotesize } , outer sep = -3.0 ] (C12) edge node {} (V12);
	\node at (2 * \dx + \dxl * 0, 40) [circle, draw, fill = \cc] (C13) {\cs};
	\node at (2 * \dx + \dxl * 0, 0) [circle, draw, fill = \vc] (V13) {\vs};
	\path [ - ,every node/.style = { font = \footnotesize } , outer sep = -3.0 ] (C13) edge node {} (V13);
	\foreach \i in {0, 1,..., 7}
	\node at (115, 70 - 10 * \i) [circle, draw, fill = \vc] (X\i) {};
	\path [ - ,every node/.style = { font = \footnotesize } , outer sep = -3.0 ]
	(U0) edge node {} (C00) (C00) edge node {} (C10) (C10) edge node {} (C30) (C30) edge node {} (X0)
	(U1) edge node {} (V00) (V00) edge node {} (C11) (C11) edge node {} (C31) (C31) edge node {} (X1)
	(U2) edge node {} (C01) (C01) edge node {} (C12) (C12) edge node {} (V30) (V30) edge node {} (X2)
	(U3) edge node {} (V01) (V01) edge node {} (C13) (C13) edge node {} (V31) (V31) edge node {} (X3)
	(U4) edge node {} (C02) (C02) edge node {} (V10) (V10) edge node {} (C32) (C32) edge node {} (X4)
	(U5) edge node {} (V02) (V02) edge node {} (V11) (V11) edge node {} (C33) (C33) edge node {} (X5)
	(U6) edge node {} (C03) (C03) edge node {} (V12) (V12) edge node {} (V32) (V32) edge node {} (X6)
	(U7) edge node {} (V03) (V03) edge node {} (V13) (V13) edge node {} (V33) (V33) edge node {} (X7);
	\node at (5, 75) {0.32};
	\node at (5, 65) {-1.33};
	\node at (5, 55) {0.99};
	\node at (5, 45) {2.51};
	\node at (5, 35) {-1.13};
	\node at (5, 25) {-1.02};
	\node at (5, 15) {-7.57};
	\node at (5, 5) {-15.53};
	
	\node at (110, 75) {-3.42};
	\node at (110, 65) {2.97};
	\node at (110, 55) {3.16};
	\node at (110, 45) {1.45};
	\node at (110, 35) {1.01};
	\node at (110, 25) {0.32};
	\node at (110, 15) {2.00};
	\node at (110, 5) {-6.12};

	\node at (20, -8) {0};
	\node at (57, -8) {2};
	\node at (97, -8) {1};
	
	\node at (122, 70) {0};
	\node at (122, 60) {1};
	\node at (122, 50) {2};
	\node at (122, 40) {3};
	\node at (122, 30) {4};
	\node at (122, 20) {5};
	\node at (122, 10) {6};
	\node at (122, 0) {7};
	
	\node at (-10, 70) {0};
	\node at (-10, 60) {1};
	\node at (-10, 50) {2};
	\node at (-10, 40) {3};
	\node at (-10, 30) {4};
	\node at (-10, 20) {5};
	\node at (-10, 10) {6};
	\node at (-10, 0) {7};
	\end{tikzpicture}
	\caption{}
	\label{FactorGraphPermutation}
\end{subfigure}
\caption{Example of the factor graph layers permutation and its influence on SC decoding. Consider that the information vector $\textbf{x} = \left[0,1,1,1\right]$ is transmitted and the vector of LLRs  $\textbf{y} = \left[-3.42, 2.97, 3.16, 1.45, 1.01, 0.32, 2.00, -6.12\right]$ is received. $\mathcal{F} = \left\lbrace 0,1,2,4 \right\rbrace $.
(\subref{PlainSC}) The plain SC decoder makes a wrong decision while processing the first information bit (highlighted in red). (\subref{LLRsPermutation}) The SC decoder returns correct codeword after applying permutation $\pi = (0,1,4,5,2,3,6,7)$ to the input LLRs vector and $\pi^{-1}$ to the decoded bits. Note that the order in which the information bits are decoded is changed. (\subref{FactorGraphPermutation}) The permutation of the factor graph layers corresponding to the permutation $\pi$ is applied.}
\label{FactorGraphPermutationExample}
\end{figure}

\section{Simulation results}
\begin{figure}[t]
\centering
\begin{tikzpicture}[ scale = 1.0 ]
	\begin{semilogyaxis}[xlabel=SNR, ylabel=BLER, xmin=-2, xmax=5, ymax=1, ymin=0.0001, legend entries={SC, SCL, Perm. decoder; Random,Perm. decoder; Dist. = 0, Perm. decoder; Dist. = 5}, grid=major, yminorgrids=true,legend pos=south west, legend cell align=left]	
	\addplot coordinates {
(-2.000000,1.000000)
(-1.750000,1.000000)
(-1.500000,0.996016)
(-1.250000,0.992063)
(-1.000000,0.992063)
(-0.750000,0.984252)
(-0.500000,0.954198)
(-0.250000,0.915751)
(0.000000,0.889680)
(0.250000,0.833333)
(0.500000,0.739645)
(0.750000,0.645995)
(1.000000,0.529661)
(1.250000,0.415282)
(1.500000,0.316456)
(1.750000,0.241779)
(2.000000,0.160462)
(2.250000,0.095602)
(2.500000,0.057326)
(2.750000,0.031447)
(3.000000,0.017663)
(3.250000,0.009138)
(3.500000,0.003654)
(3.750000,0.001617)
(4.000000,0.000788)
(4.250000,0.000273)
(4.500000,0.000124)
(4.750000,0.000062)
};
\addplot coordinates {
(-2.000000,1.000000)
(-1.750000,0.996016)
(-1.500000,0.988142)
(-1.250000,0.984252)
(-1.000000,0.957854)
(-0.750000,0.899281)
(-0.500000,0.838926)
(-0.250000,0.755287)
(0.000000,0.675676)
(0.250000,0.577367)
(0.500000,0.437828)
(0.750000,0.316857)
(1.000000,0.219491)
(1.250000,0.154991)
(1.500000,0.090481)
(1.750000,0.056016)
(2.000000,0.032753)
(2.250000,0.019042)
(2.500000,0.011621)
(2.750000,0.006590)
(3.000000,0.003912)
(3.250000,0.002009)
(3.500000,0.001126)
(3.750000,0.000714)
(4.000000,0.000410)
(4.250000,0.000185)
(4.500000,0.000105)
(4.750000,0.000053)
};
\addplot coordinates {
(-3.000000,1.000000)
(-2.750000,1.000000)
(-2.500000,1.000000)
(-2.250000,1.000000)
(-2.000000,0.996016)
(-1.750000,0.996016)
(-1.500000,0.992063)
(-1.250000,0.988142)
(-1.000000,0.984252)
(-0.750000,0.976562)
(-0.500000,0.950570)
(-0.250000,0.909091)
(0.000000,0.880282)
(0.250000,0.811688)
(0.500000,0.710227)
(0.750000,0.609756)
(1.000000,0.486381)
(1.250000,0.373692)
(1.500000,0.269107)
(1.750000,0.185185)
(2.000000,0.113225)
(2.250000,0.063906)
(2.500000,0.032757)
(2.750000,0.016262)
(3.000000,0.007644)
(3.250000,0.003519)
(3.500000,0.001514)
(3.750000,0.000832)
(4.000000,0.000430)
(4.250000,0.000192)
(4.500000,0.000107)
(4.750000,0.000054)
};
\addplot coordinates {
(-2.000000,1.000000)
(-1.750000,1.000000)
(-1.500000,0.988142)
(-1.250000,0.988142)
(-1.000000,0.976562)
(-0.750000,0.965251)
(-0.500000,0.909091)
(-0.250000,0.868056)
(0.000000,0.811688)
(0.250000,0.718391)
(0.500000,0.623441)
(0.750000,0.499002)
(1.000000,0.395570)
(1.250000,0.291375)
(1.500000,0.201939)
(1.750000,0.137589)
(2.000000,0.080515)
(2.250000,0.044587)
(2.500000,0.025184)
(2.750000,0.015387)
(3.000000,0.007673)
(3.250000,0.003717)
(3.500000,0.001761)
(3.750000,0.000944)
(4.000000,0.000495)
(4.250000,0.000215)
(4.500000,0.000111)
(4.750000,0.000058)
};
\addplot coordinates {
(-2.000000,1.000000)
(-1.750000,1.000000)
(-1.500000,0.992063)
(-1.250000,0.992063)
(-1.000000,0.984252)
(-0.750000,0.965251)
(-0.500000,0.912409)
(-0.250000,0.853242)
(0.000000,0.755287)
(0.250000,0.672043)
(0.500000,0.584112)
(0.750000,0.428082)
(1.000000,0.309406)
(1.250000,0.215332)
(1.500000,0.141965)
(1.750000,0.085470)
(2.000000,0.045704)
(2.250000,0.025757)
(2.500000,0.014309)
(2.750000,0.007391)
(3.000000,0.004357)
(3.250000,0.002133)
(3.500000,0.001177)
(3.750000,0.000718)
(4.000000,0.000414)
(4.250000,0.000186)
(4.500000,0.000105)
(4.750000,0.000053)
};
	\end{semilogyaxis}
	\end{tikzpicture}
	\caption{The error correction performance of the polar code of
dimension 128 and length 256. List size equals 16.}
\label{perms_effects}
\end{figure}
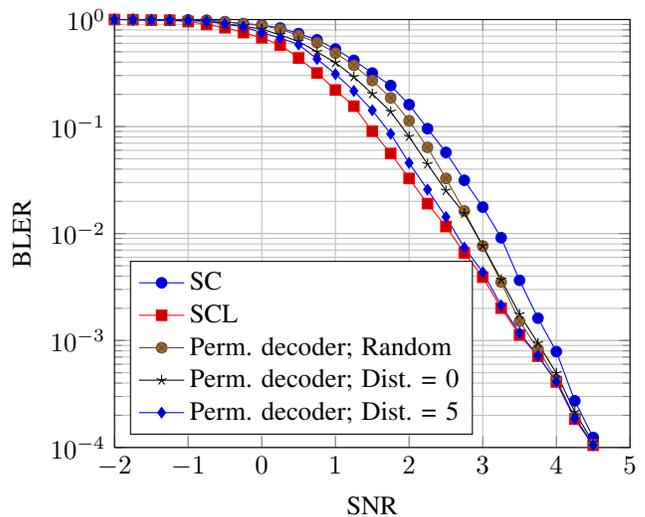

It is assumed that the transmission is performed over a BI-AWGN channel and block error rate (BLER) is estimated. Effect of different permutations sets on the error correction performance of the considered permutation decoder is depicted in Fig. \ref{perms_effects}. The polar code of length 256 and dimension 128 obtained using the construction adopted by the 3rd Generation Partnership Project (3GPP) \cite{3GPP} is considered. The list size equals 16. A random permutations set contains trivial permutation and 15 random ones. Also, permutations sets optimized using (\ref{pscl_lower_bound}) with the minimum Hamming distance between the elements constraint are taken into account. Simulation results demonstrate that the considered permutation decoding algorithm with the permutations set optimized with the minimum Hamming distance constraint allows getting the error correction performance close to that of the SCL decoder. Namely, performance degradation is at most 0.25 dB. The random permutations set, as well as the set optimized without minimum Hamming distance constraint, demonstrates performance degradation at least 0.5 dB for high Signal-to-Noise Ratio (SNR) region.

\begin{figure}[t]
\centering
\begin{tikzpicture}[ scale = 1.0 ]
	\begin{semilogyaxis}[xlabel=SNR, ylabel=BLER, xmin=-4, xmax=2.25, ymax=1, ymin=0.0001, legend entries={Fixed; SCL., Fixed; Perm. decoder., Joint; SCL., Joint; Perm. decoder.}, grid=major, yminorgrids=true,legend pos=south west, legend cell align=left]	
	\addplot coordinates {
(-4.000000,1.000000)
(-3.750000,1.000000)
(-3.500000,0.996016)
(-3.250000,0.976562)
(-3.000000,0.936330)
(-2.750000,0.892857)
(-2.500000,0.814332)
(-2.250000,0.683060)
(-2.000000,0.550661)
(-1.750000,0.423729)
(-1.500000,0.283447)
(-1.250000,0.174459)
(-1.000000,0.110035)
(-0.750000,0.068757)
(-0.500000,0.041813)
(-0.250000,0.022808)
(0.000000,0.013363)
(0.250000,0.006705)
(0.500000,0.003858)
(0.750000,0.002118)
(1.000000,0.001277)
(1.250000,0.000700)
(1.500000,0.000353)
(1.750000,0.000181)
(2.000000,0.000088)
};
\addplot coordinates {
(-4.000000,1.000000)
(-3.750000,1.000000)
(-3.500000,0.992063)
(-3.250000,0.992063)
(-3.000000,0.988142)
(-2.750000,0.961538)
(-2.500000,0.922509)
(-2.250000,0.853242)
(-2.000000,0.759878)
(-1.750000,0.644330)
(-1.500000,0.483559)
(-1.250000,0.355114)
(-1.000000,0.229991)
(-0.750000,0.143843)
(-0.500000,0.085646)
(-0.250000,0.042130)
(0.000000,0.020706)
(0.250000,0.009269)
(0.500000,0.004698)
(0.750000,0.002415)
(1.000000,0.001352)
(1.250000,0.000742)
(1.500000,0.000358)
(1.750000,0.000184)
(2.000000,0.000089)
};
\addplot coordinates {
(-3.000000,1.000000)
(-2.750000,1.000000)
(-2.500000,0.984252)
(-2.250000,0.965251)
(-2.000000,0.932836)
(-1.750000,0.859107)
(-1.500000,0.778816)
(-1.250000,0.634518)
(-1.000000,0.520833)
(-0.750000,0.409836)
(-0.500000,0.272035)
(-0.250000,0.174948)
(0.000000,0.105843)
(0.250000,0.051856)
(0.500000,0.023614)
(0.750000,0.010258)
(1.000000,0.003383)
(1.250000,0.001051)
(1.500000,0.000253)
(1.750000,0.000052)
};
\addplot coordinates {
(-3.000000,1.000000)
(-2.750000,1.000000)
(-2.500000,0.988142)
(-2.250000,0.972763)
(-2.000000,0.932836)
(-1.750000,0.877193)
(-1.500000,0.791139)
(-1.250000,0.683060)
(-1.000000,0.560538)
(-0.750000,0.414594)
(-0.500000,0.266525)
(-0.250000,0.166667)
(0.000000,0.089606)
(0.250000,0.041904)
(0.500000,0.018699)
(0.750000,0.006023)
(1.000000,0.001912)
(1.250000,0.000543)
(1.500000,0.000153)
(1.750000,0.000034)
};
	\end{semilogyaxis}
	\end{tikzpicture}
	\caption{The error correction performance of  polar code of dimension 170 and length 512. List size equals 16.}
\label{polar_170}
\end{figure}
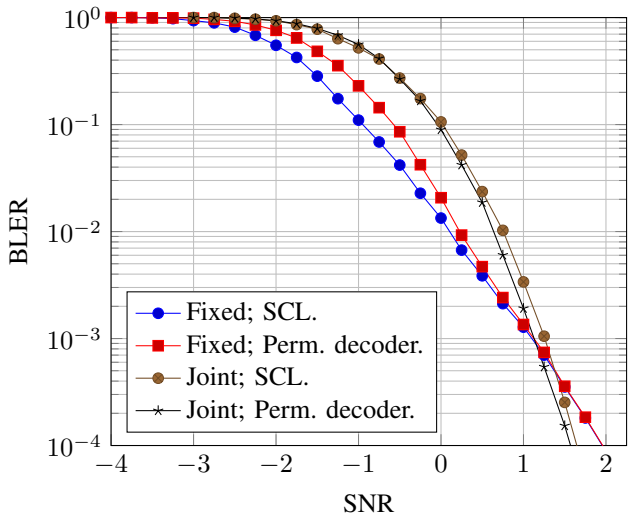

\begin{figure}[th]
\centering
\begin{tikzpicture}[ scale = 1.0 ]
	\begin{semilogyaxis}[xlabel=SNR, ylabel=BLER, xmin=-2, xmax=4.75, ymax=1, ymin=0.0001, legend entries={Fixed; SCL., Fixed; Perm. decoder., Joint; SCL., Joint; Perm. decoder.}, grid=major, yminorgrids=true,legend pos=south west, legend cell align=left]	
	\addplot coordinates {
(-2.000000,1.000000)
(-1.750000,1.000000)
(-1.500000,1.000000)
(-1.250000,0.996016)
(-1.000000,0.996016)
(-0.750000,0.988142)
(-0.500000,0.965251)
(-0.250000,0.932836)
(0.000000,0.801282)
(0.250000,0.668449)
(0.500000,0.541126)
(0.750000,0.391850)
(1.000000,0.259067)
(1.250000,0.152905)
(1.500000,0.095129)
(1.750000,0.053259)
(2.000000,0.030425)
(2.250000,0.017442)
(2.500000,0.008777)
(2.750000,0.004651)
(3.000000,0.002788)
(3.250000,0.001610)
(3.500000,0.000842)
(3.750000,0.000431)
(4.000000,0.000225)
(4.250000,0.000119)
(4.500000,0.000057)
};
\addplot coordinates {
(-1.000000,1.000000)
(-0.750000,0.996016)
(-0.500000,0.992063)
(-0.250000,0.992063)
(0.000000,0.946970)
(0.250000,0.836120)
(0.500000,0.722543)
(0.750000,0.568182)
(1.000000,0.433276)
(1.250000,0.286369)
(1.500000,0.174581)
(1.750000,0.093423)
(2.000000,0.043875)
(2.250000,0.024366)
(2.500000,0.011061)
(2.750000,0.005284)
(3.000000,0.003053)
(3.250000,0.001650)
(3.500000,0.000847)
(3.750000,0.000435)
(4.000000,0.000228)
(4.250000,0.000121)
(4.500000,0.000057)
};
\addplot coordinates {
(-2.000000,1.000000)
(-1.750000,1.000000)
(-1.500000,1.000000)
(-1.250000,1.000000)
(-1.000000,0.996016)
(-0.750000,0.996016)
(-0.500000,0.996016)
(-0.250000,0.996016)
(0.000000,0.984252)
(0.250000,0.965251)
(0.500000,0.905797)
(0.750000,0.811688)
(1.000000,0.700280)
(1.250000,0.523013)
(1.500000,0.367107)
(1.750000,0.231267)
(2.000000,0.132767)
(2.250000,0.063955)
(2.500000,0.028982)
(2.750000,0.010102)
(3.000000,0.002588)
(3.250000,0.000677)
(3.500000,0.000150)
(3.750000,0.000035)
};
\addplot coordinates {
(-1.000000,1.000000)
(-0.750000,1.000000)
(-0.500000,0.996016)
(-0.250000,0.992063)
(0.000000,0.988142)
(0.250000,0.965251)
(0.500000,0.943396)
(0.750000,0.862069)
(1.000000,0.720461)
(1.250000,0.565611)
(1.500000,0.403877)
(1.750000,0.245098)
(2.000000,0.134844)
(2.250000,0.059880)
(2.500000,0.022541)
(2.750000,0.007351)
(3.000000,0.002110)
(3.250000,0.000624)
(3.500000,0.000145)
(3.750000,0.000033)
};
	\end{semilogyaxis}
	\end{tikzpicture}
	\caption{The error correction performance of polar code of dimension 256 and length 512. List size equals 16.}
\label{polar_256}
\end{figure}
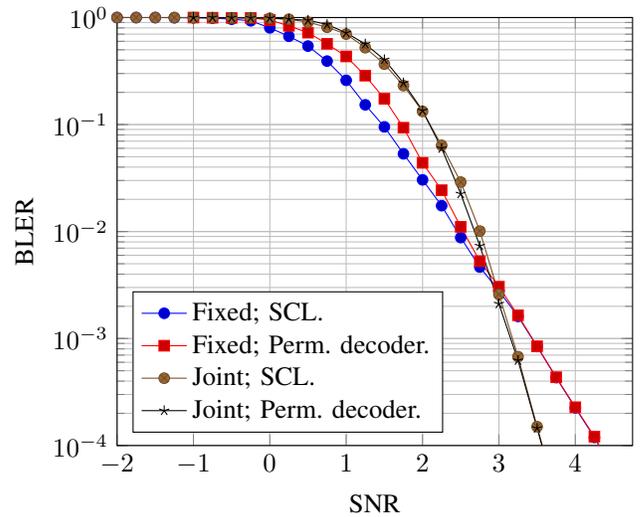

The error correction performance of polar codes of length 512 and dimensions 170 and 256 under the SCL and the permutation decoding algorithm with list size 16 are presented in Fig. \ref{polar_170} and Fig. \ref{polar_256} respectively. For the fixed frozen bits set scenario, indices of frozen subchannels are selected in accordance with the procedure adopted by the 3GPP \cite{3GPP}, while the permutations set is optimized using lower bound for the error correction performance (\ref{pscl_lower_bound}). Moreover, it is considered that Hamming distance between any two permutations in the permutations set is greater or equal than 5. For joint optimization scenario, the set of frozen bits is optimized using GA for density evolution, while the permutations set is a subset of all available permutations $\left\lbrace\pi^l : \pi^l\left(0\right) = 0, \pi^l\left(1\right) = 1,  \pi^l\left(2\right) = 2,  \pi^l\left(3\right) = 3\right\rbrace$. Thus, the maximum list size equals $120$.

In case of the predefined frozen bits set, the permutation
decoder demonstrates error correction performance degradation up to 0.25 dB, while the joint optimization of both the frozen bits set and the permutations set allows getting codes having the same error correction performance under the SCL and the permutation decoding algorithms. Moreover, joint approach outperforms 3GPP polar codes in high SNR region under the proposed permutation decoding method for the considered dimensions.

\section{Conclusions}
A novel permutation decoding approach for polar codes is presented. The main benefit of the proposed method is absence of sorting operations making it more feasible for hardware implementation in comparison with the SCL algorithm. It is shown that it is possible to construct polar codes having similar performance under the SCL and considered permutation decoding while outperforming state-of-the-art 3GPP polar codes in high SNR region under SCL decoding with list size 16. Moreover, it is demonstrated that it is possible to optimize a permutations set for the fixed frozen bits set, with the error correction performance degradation being at most 0.25 dB. The construction of the permutations set is based on minimization of the lower bound for the error correction performance, while the exact theoretical analysis of the block error probability is an open problem.

\bibliographystyle{IEEEtran}
\bibliography{IEEEabrv,myBib}

\begin{thebibliography}{10}
\providecommand{\url}[1]{#1}
\csname url@samestyle\endcsname
\providecommand{\newblock}{\relax}
\providecommand{\bibinfo}[2]{#2}
\providecommand{\BIBentrySTDinterwordspacing}{\spaceskip=0pt\relax}
\providecommand{\BIBentryALTinterwordstretchfactor}{4}
\providecommand{\BIBentryALTinterwordspacing}{\spaceskip=\fontdimen2\font plus
\BIBentryALTinterwordstretchfactor\fontdimen3\font minus
  \fontdimen4\font\relax}
\providecommand{\BIBforeignlanguage}[2]{{%
\expandafter\ifx\csname l@#1\endcsname\relax
\typeout{** WARNING: IEEEtran.bst: No hyphenation pattern has been}%
\typeout{** loaded for the language `#1'. Using the pattern for}%
\typeout{** the default language instead.}%
\else
\language=\csname l@#1\endcsname
\fi
#2}}
\providecommand{\BIBdecl}{\relax}
\BIBdecl

\bibitem{Arikan}
E.~Arikan, ``Channel polarization: A method for constructing capacity-achieving
  codes for symmetric binary-input memoryless channels,'' \emph{IEEE
  Transactions on Information Theory}, vol.~55, no.~7, pp. 3051--3073, July
  2009.

\bibitem{SCL}
I.~Tal and A.~Vardy, ``List decoding of polar codes,'' \emph{IEEE Transactions
  on Information Theory}, vol.~61, no.~5, pp. 2213--2226, May 2015.

\bibitem{MinSum}
A.~Balatsoukas-Stimming, M.~B. Parizi, and A.~Burg, ``{LLR}-based successive
  cancellation list decoding of polar codes,'' \emph{IEEE Transactions on
  Signal Processing}, vol.~63, no.~19, pp. 5165--5179, Oct 2015.

\bibitem{SCFlip}
O.~Afisiadis, A.~Balatsoukas-Stimming, and A.~Burg, ``A low-complexity improved
  successive cancellation decoder for polar codes,'' in \emph{2014 48th
  Asilomar Conference on Signals, Systems and Computers}, Nov 2014, pp.
  2116--2120.

\bibitem{SCFlip2}
L.~Chandesris, V.~Savin, and D.~Declercq, ``Dynamic-{SCF}lip decoding of polar
  codes,'' \emph{IEEE Transactions on Communications}, vol.~66, no.~6, pp.
  2333--2345, June 2018.

\bibitem{GA}
P.~Trifonov, ``Efficient design and decoding of polar codes,'' \emph{IEEE
  Transactions on Communications}, vol.~60, no.~11, pp. 3221--3227, November
  2012.

\bibitem{Sloan}
F.~J. MacWilliams and N.~J.~A. Sloane, \emph{The theory of error-correcting
  codes}.\hskip 1em plus 0.5em minus 0.4em\relax Amsterdam, The Netherlands:
  North-Holland, 1977.

\bibitem{PermGroupPolar}
M.~Bardet, V.~Dragoi, A.~Otmani, and J.~Tillich, ``Algebraic properties of
  polar codes from a new polynomial formalism,'' in \emph{2016 IEEE
  International Symposium on Information Theory (ISIT)}, July 2016, pp.
  230--234.

\bibitem{Perm}
\BIBentryALTinterwordspacing
N.~Doan, S.~A. Hashemi, M.~Mondelli, and W.~J. Gross, ``On the decoding of
  polar codes on permuted factor graphs.'' [Online]. Available:
  \url{https://arxiv.org/abs/1806.11195v1}
\BIBentrySTDinterwordspacing

\bibitem{PermSC}
H.~Vangala, E.~Viterbo, and Y.~Hong, ``Permuted successive cancellation decoder
  for polar codes,'' in \emph{2014 International Symposium on Information
  Theory and its Applications}, Oct 2014, pp. 438--442.

\bibitem{SCHardware}
C.~Leroux, I.~Tal, A.~Vardy, and W.~J. Gross, ``Hardware architectures for
  successive cancellation decoding of polar codes,'' in \emph{2011 IEEE
  International Conference on Acoustics, Speech and Signal Processing
  (ICASSP)}, May 2011, pp. 1665--1668.

\bibitem{RMPerm}
\BIBentryALTinterwordspacing
M.~Kamenev, Y.~Kameneva, O.~Kurmaev, and A.~Maevskiy, ``A new permutation
  decoding method for {R}eed-{M}uller codes.'' [Online]. Available:
  \url{https://arxiv.org/abs/1901.04433}
\BIBentrySTDinterwordspacing

\bibitem{3GPP}
\BIBentryALTinterwordspacing
\emph{3GPP TS 38.212 Multiplexing and channel coding}, Release 15.2.0, 2018.
  [Online]. Available:
  \url{http://3gpp.org/ftp/specs/2018-06/rel-15/38\_series/38212-f20.zip}
\BIBentrySTDinterwordspacing

\end{thebibliography}

\end{document}